# Raman laser from an optical resonator with a grafted single molecule monolayer

*Xiaoqin Shen[1, *, ‡], Hyungwoo Choi[1,*], Dongyu Chen[2], Wei Zhao[3], and Andrea M. Armani[1,2]*

[1]Mork Family Department of Chemical Engineering and Material Science, University of Southern California, Los Angeles, California, 90089, United States

[2]Ming Hsieh Department of Electrical Engineering-Electrophysics, University of Southern California, Los Angeles, California, 90089, United States

[3]Department of Chemistry, University of Arkansas at Little Rock, Little Rock, Arkansas, 72204, United States

*The authors contributed equally to this work.

‡Currently located at School of Physical Science and Technology, ShanghaiTech University, Shanghai, China.

Abstract: Raman-based technologies have enabled many ground-breaking scientific discoveries related to surface science, single molecule chemistry and biology. For example, researchers have identified surface bound molecules by their Raman vibrational modes and demonstrated polarization-dependent Raman gain. However, a surface constrained Raman laser has yet to be demonstrated because of the challenges associated with achieving a sufficiently high photon population located at a surface to transition from spontaneous to stimulated Raman scattering. Here, advances in surface chemistry and in integrated photonics are combined to demonstrate lasing based on surface stimulated Raman scattering (SSRS). By creating an oriented, constrained Si-O-Si monolayer on the surface of integrated silica optical microresonators, the requisite conditions for SSRS are achieved with low threshold powers (200μW). The expected polarization-dependence of the SSRS due to the orientation of the Si-O-Si bond is observed. Due to the ordered monolayer, the Raman lasing efficiency is improved from ~5% to over 40%.

Introduction

Since the first Raman laser based on a gas cell[1], Raman lasers have been demonstrated using a wide range of platforms including optical fibers[2–4], integrated devices[5–7], nanocrystalline particles[8–10], and bulk crystals[11,12]. These lasers rely on a nonlinear optical process called stimulated Raman scattering[13]. In a spontaneous Raman scattering process, an initial pump light interacts with a nonlinear medium and spontaneously generates a longer wavelength emission signal. This signal is Stokes-shifted from the initial pump wavelength by a frequency difference equal to the atomic or molecular vibration frequency of the material[14]. If a sufficient population of Stokes-shifted photons is present when a new Stokes-shifted photon is generated, the rate of the photon generation is increased, and the system transitions from the linear spontaneous Raman to the nonlinear stimulated Raman scattering (SRS) process[13]. Due to the versatility and tunability of the emission wavelength, leveraging SRS for laser development is a popular strategy. However, these devices all rely on the bulk material for generation of the SRS which is distinctly different from a surface SRS (surface stimulated Raman scattering) behavior[15–17].

In a surface Raman process, the Raman scattering intensity is dominated by the orientation of the vibrational mode with respect to the polarization of the incident wave at the surface[18–20]. Therefore, achieving SSRS requires two criteria to be met simultaneously. First, the polarization of the incident optical field must be aligned with the vibrational mode of the molecule[13,18], and second, there must be a sufficient number of incident photons to transition from spontaneous to stimulated Raman scattering. Therefore, one key challenge is efficiently exciting or injecting photons into the surface layer. Previous work has investigated Raman signals generated by molecular monolayers deposited randomly on the surface[15]; however, only extremely weak Raman signals were observed in these measurements due to the lack of alignment of the Raman gain with the incident field[21]. Additionally, given the random nature of these films, all information regarding the polarization at a molecular level was obscured. Therefore, to realize SSRS and study the fundamental physics, it is necessary to establish a high intensity optical field of a single polarization that is aligned with the molecular vibrational modes. This precision requires a combination of innovations in optical physics and surface chemistry.

One solution for creating the required highly defined and intense optical source can be found in integrated photonics[22]. Many on-chip devices including waveguides and resonators support evanescent fields that create the requisite surface propagating excitation source (Figure 1a). Among the possible device types, whispering gallery mode optical cavities are particularly unique. These devices confine light at well-defined resonant wavelengths in circular orbits at the device-air interface with 1-5% of the optical field extending just beyond the cavity surface depending on the precise device geometry and material properties[22]. In addition, whispering gallery mode resonators with functional coatings can have quality factors (Q) in excess of 10 million, resulting in long photon lifetimes. As a result, they can act as optical amplifiers, enabling a build-up of optical power that can facilitate nonlinear phenomena[23–28]. For example, a 50 µm diameter silica toroidal resonator with a Q of 10 million in the near-IR can increase an input power of 1 mW to approximately 100 W. In previous work, these ultra-high Q factors have been leveraged to achieve µW Raman lasing thresholds in silica devices despite the low Raman gain[29,30].

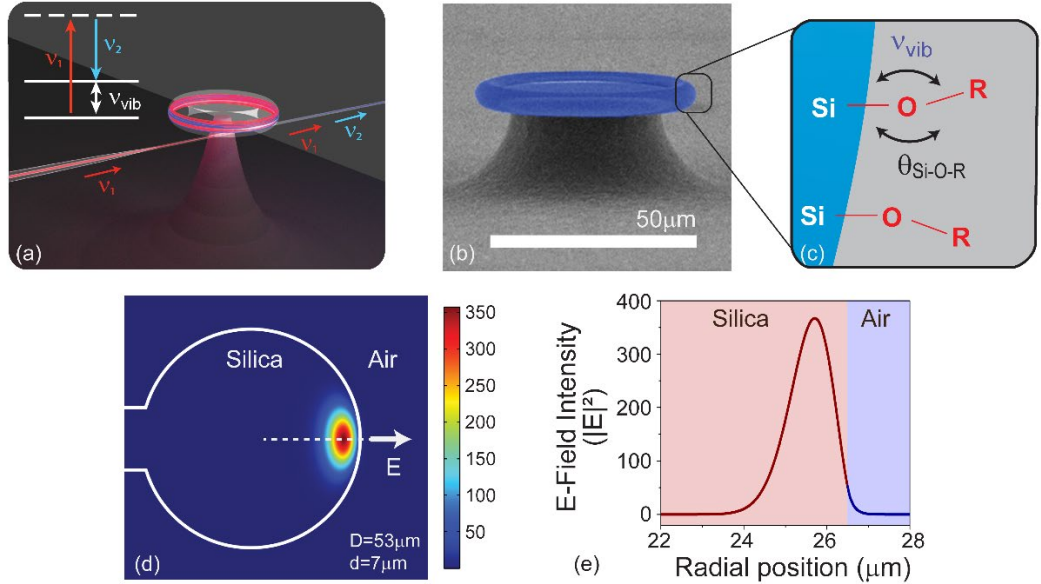

Figure 1. Excitation of SSRS using integrated optical microcavity. (a) Rendering of the silica toroidal microcavity laser. A waveguide couples pump light ($v_1$) into and Raman emission ($v_2$) out of the cavity. Inset: $v_2$ is related to the vibrational frequency ($v_{vib}$) of the Si-O-Si mode being excited. (b) Scanning electron microscope (SEM) image of a silica microtoroidal resonator. False blue colour indicates the grafted siloxane molecular layer. (c) Schematic of the cavity surface showing the tilted surface Raman modes (Si-O-R), where R can be any residue. (d) Finite element method (FEM) simulation of the optical mode profile in a microtoroidal resonator. The white arrow represents the electric field direction of the fundamental transverse magnetic (TM) mode. (e) The optical mode profile as indicated by the dashed white line in (d). The shaded blue region is able to interact with the surface-bound monolayer.

In addition, ultra-high-Q resonators overcome the challenge associated with selective probing of the polarization of the molecular Raman mode. For directional Raman modes, such as those generated by whispering gallery mode optical cavities, the Raman scattering intensity or efficiency (*S*) is dominated by the mode orientation (Raman tensor) and polarization geometry. The relation can be expressed as $S \propto |e_s \cdot \alpha_R \cdot e_i|^2$, where $e_s$ and $e_i$ are the polarization vector of the scattered and incident light respectively, and $\alpha_R$ is the Raman tensor of a specific vibrational mode[32,33]. The theoretical angular dependent Raman intensity will vary as $\cos^2(\theta)$, where $\theta$ is the angle between the direction of the Raman mode and the electrical field[20,31] (Figure 1b, c). This expression suggests that the stimulated Raman scattering efficiency for a Raman mode oriented perpendicular to the surface is maximized when the mode orientation is aligned with the direction of the electric field. Silica optical resonators support both TE and TM modes which are located at distinct optical frequencies. Therefore, by selectively exciting the individual cavity modes, the polarization dependence of the surface Raman behaviour can be studied with exquisite sensitivity. However, the alignment of the vibrational frequency ($\nu_{vib}$) with the optical field is critical in these measurements. Therefore, a high precision surface chemistry that does not degrade the optical performance of the optical resonator and that ensures this orientation must be implemented.

In the present work, we combine surface chemistry and integrated photonics to demonstrate surface stimulated Raman scattering (SSRS) emission, achieving a surface Raman laser. Organic siloxane single-molecular layers are grafted onto the surface of a toroidal optical microresonator, resulting in a highly oriented Si-O-Si vibrational surface Raman mode. The vibrational Si-O-Si mode is excited with the high intensity whispering gallery optical mode, and the polarization dependence of the surface Raman behaviour is investigated by leveraging either the TE or the TM optical resonant modes[7,33]. As a result of the selective interaction between the polarized optical field and the highly aligned surface Raman mode, surface Raman lasing behaviours are generated with dramatically enhanced Raman lasing performance from ~5% to over ~40% efficiency compared to bulk silica devices, and the polarization dependence of the SSRS is experimentally demonstrated in agreement with theoretical predictions.

Results and Discussion

The silica toroidal whispering gallery mode resonators integrated on silicon shown in Figure 1 (b) are used as the optical resonator[34]. Two different device sizes are investigated as part of this work: 1) major diameter of 52.7 ± 8.8 µm with minor diameter of 6.7 ± 0.6 µm, and 2) major diameter of 83.3 ± 2.1 µm with minor diameter of 11.6 ± 0.7 µm. Device geometry, both major and minor diameter, governs the lasing threshold power because the optical circulating intensity is inversely proportional to device diameter. However, efficiency should be independent of either parameter. Therefore, studying multiple sizes is informative. Figure 1 (d) contains the fundamental transverse magnetic (TM) optical mode simulated via COMSOL Multiphysics finite element method (FEM) for a 53µm diameter device. The direction of the electric field of the TM mode is along the horizontal, whereas that of transverse electric (TE) mode is perpendicular to it. Normalized electric field distribution of the fundamental TM mode in Figure 1 (e) indicates that ~5% of the optical field is at the surface of the microresonator.

Three different surface monolayer functional groups are studied (Figure 2 (a) - (c)). The first surface group investigated is the hydroxyl (OH) layer that is intrinsic to the silica device. This layer is then exchanged with an organic methylsiloxane (-MS) or dimethylsiloxane (-DMS) molecular monolayer using a chemical vapor deposition process (Figure S1). During the process, highly reactive chlorosilane molecules react with the surface hydroxyl groups based on the silanization reaction, yielding new oriented Si-O-Si bonds on surface[35,36]. The constrained Si-O-Si bond angle is between ~ 120° and ~ 125° (Figure S2 and references[37,38]). Previous work has shown that this reaction is non-destructive to the device optical behaviour[39] and intrinsically self-limiting, allowing only a single molecular monolayer to form as indicated in Figure 2. Additional verification of the surface chemistry is included in Figure S3.

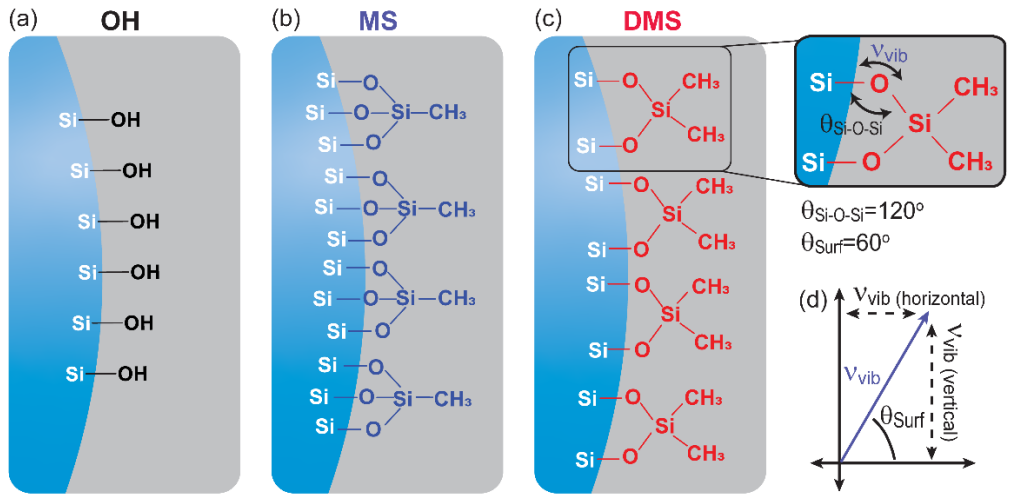

Figure 2. Schematic images of the surfaces of (a) OH, (b) MS, and (c) DMS monolayer-functionalized resonator devices. The blue indicates silica, and the gray indicates air. Zoom-in image shows the orientation of the tilted surface Raman modes (Si-O-Si) that arise due to the specific organosiloxane surface chemistry used in the present work. These vibrational modes are only present in the DMS and MS functionalized devices. (d) Vector decomposition of the horizontal and vertical components of the Si-O-Si mode with respect to the toroidal cavity surface.

Given this combination of materials, possible Raman signatures ($\nu_{vib}$) could be generated by the oriented Si-O-Si or the C-H chemical bonds (Figure 2). Assuming the Si-O-Si bond has a fixed 120º angle with respect to the surface, the overall Raman mode will be tilted, theoretically with an angle of ~ 60º with respect to the cavity surface or ~ 30º off-axis (Figure 2 (d)). Therefore, the effect of the surface layer will influence both the TM and TE polarizations to differing degrees. Specifically, based on these angles, the ratio of Raman efficiencies between TM and TE modes can be approximately calculated to be $\cos^2(30°)/\cos^2(60°) = 3$. If the bond angle is slightly larger (125°), this ratio will increase to 3.69. In contrast, the C-H chemical bond angle is not fixed with respect to the resonator surface. Additional details on device fabrication and modelling, surface functionalization chemistry, and control measurements are in the Supporting Information.

To complement the measurement with the optical resonators, several theoretical and experimental control studies are performed. Density functional theory (DFT) simulations of model molecules related to those used in the experimental work are used to calculate the vibrational second hyperpolarizabilities ($\gamma^{vib}$) [40]. The $\gamma^{vib}$ corresponding to the Si-O-Si mode of the individual molecules are calculated to be 1.2 and 1.7 x $10^{-36}$ cm$^6$/erg for the MS and DMS model molecules, respectively (Figure S4 and Table S1). These values are about two times larger than the silicon dioxide model molecule (0.78 x $10^{-36}$ cm$^6$/erg) due to the asymmetry that is introduced by the organic methyl group in the siloxane structure. In addition, by spin-casting each material on silicon wafers, thin film samples are prepared and analysed with Raman spectroscopy (Figure S2). While this method is limited to the detection of spontaneous Raman signals, it allows for the identification of the Si-O-Si vibration. In the MS and DMS samples, this peak is clearly identifiable at about 460 cm$^{-1}$, as expected. The gain coefficient values for the MS and DMS layers are calculated to be ~13.5 × $10^{-13}$ m/W and ~12.9 × $10^{-13}$ m/W, with silica (0.53 × $10^{-13}$ m/W) as reference[41]. The higher Raman gain values of MS and DMS molecules than $SiO_2$ are consistent with the data from the DFT simulation results. However, while informative, it is important to note that the DFT studies are of single molecules in isolation, and the Raman measurements are of disordered films, not molecular monolayers anchored to a surface. Details on the DFT model, Raman spectra measurements, and gain calculation are in the SI.

To characterize the optical resonator quality factor (Q) and the quantify the SSRS behaviour, the analysis set-up in Figure 3(a) is used. Light from a 765nm tunable laser is coupled into the cavity using the single mode tapered fibre waveguide. A resonant optical wavelength ($\lambda_o$) of the cavity is identified, and the intrinsic cavity Q is determined by measuring the linewidth of the resonance over a range of coupling conditions (Figure S5)[42]. The Raman emission characteristics are recorded on an optical spectrum analyser (OSA) while simultaneously monitoring the cavity Q. The TE and the TM mode of the optical cavity are selectively excited by changing the polarization of the input optical light. The lasing threshold and slope efficiency are determined by analyzing the change in first order Raman lasing power as a function of the coupled power and the polarization state[43]. Additional details and all data are in the SI (Table S2 and S3).

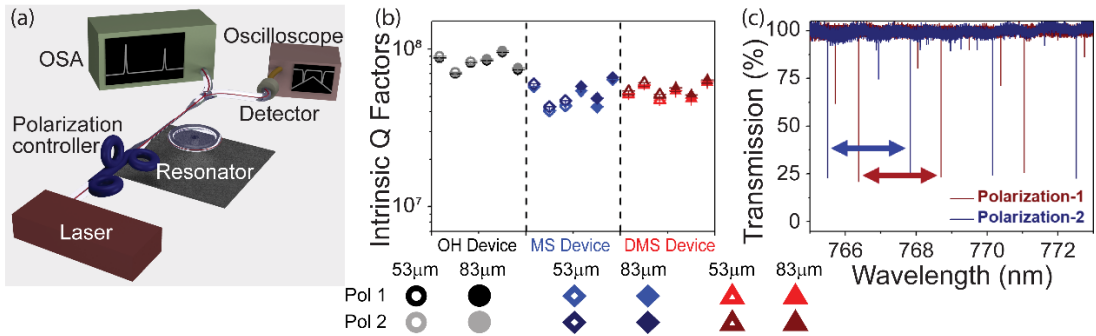

Figure 3. Device characterization method and optical device performance. (a) Schematic image of the device characterization set-up. Light from a tunable laser centered at 765 nm is coupled into the optical resonator using a tapered optical fibre waveguide. The polarization state of the input light is controlled by polarization controller. The output signal from the resonator is split using a 50/50 fibre splitter with one output going to an optical spectrum analyser (OSA) and the other connecting to a photodetector. (b) The intrinsic Q factors of a series of -OH, -MS, and -DMS devices are measured. Two different device sizes (~53 µm (hollow symbol) and ~83 µm (solid symbol)) are characterized at two different polarization states. Error bars are smaller than symbols. (c) One of the representative broad-scan spectra of a MS-functionalized device with two different polarization states (blue and red). Additional details are contained in the SI.

An important feature of optical cavities is that the resonant wavelength ($\lambda_o$) is governed, in part, by the cavity geometry; thus, the resonant wavelength is unique to the cavity. Moreover, Raman can only be generated when a cavity is on-resonance. As such, the resonant wavelength and the pump wavelength ($\lambda_{pump}$) are identical ($\lambda_o = \lambda_{pump}$). However, the Raman shift, or the difference between the pump frequency ($\upsilon_{pump}$) and Raman emission frequency ($\upsilon_{Raman}$), is governed by the vibrational mode of the molecule. Therefore, when comparing data from different cavities, while the Raman emission wavelength may vary, the Raman shift should be conserved.

Two key enabling features of using whispering gallery mode cavities to study SSRS are the ultra-high-Q and the ability to selectively excite either the TE or TM optical modes. Therefore, it is critical to maintain both capabilities, post-surface functionalization. The Q factors for both the TM and TE modes are plotted in Figure 3 (b), and each pair of data points represents a unique device. All devices have ultra-high Q factors between $1\times10^7$ to $1\times10^8$, demonstrating the reproducibility of the fabrication and the negligible impact of the surface chemistry on the device's optical Q. Figure 3 (c) is a pair of broad-scan spectra from a 53 μm diameter MS-functionalized device. Using an in-line polarization controller, the TM and TE mode families are identified in the broad scan spectrum and isolated to study the polarization dependence of the SSRS behaviour. The deepest peaks from each spectrum indicate the fundamental TM or TE mode. The free spectral range (FSR) of both spectra is ~2.33 nm which is similar to the theoretical prediction of 2.4 nm.

Figure 4(a)-(c) contains representative emission spectra measured on the OSA from the three device types with ~ 350 μW of coupled power. All three devices show Raman emission when the device is on-resonance. In the -OH functionalized silica device (Figure 4 (a, d)), the Raman emission is solely from the SRS of OH device, where the Si-O-Si vibrational Raman mode is randomly distributed with respect to the optical field. In contrast, in the devices functionalized with -MS and -DMS, the surface Si-O-Si Raman mode is oriented. As a result, the Raman lasing is more pronounced. Comparing the results for the three devices (Figure 4 (d)), MS and DMS devices show a significant increase in the emission signal of about 8 times (44.71 and 50.02 µW, respectively) as compared to the OH device (6.57 µW). Across all devices studied, the Raman shift values of MS and DMS devices vary slightly from ~420 to ~500 cm$^{-1}$ as shown at Figure 4e. These values fall within the Raman gain band of the Si-O-Si vibrational mode as determined in the control measurements (Figure S2) and as published previously[44], allowing the assignment of the emission to the Si-O-Si mode. It is worth noting that emission peaks from the C-H stretching modes of the methyl groups which would occur around 980 nm (~2850 cm$^{-1}$ to ~2980 cm$^{-1}$, see Figure S2 and reference[45]) are not observed, even under high input power (Figure S6). In contrast to the oriented Si-O-Si vibrational Raman mode, the C-H stretching modes of the methyl groups in the surface molecules are randomly distributed and not ordered. Therefore, the absence of emission peaks at ~980 nm provides evidence that molecular attachment and orientation of the vibrational Raman mode with respect to the optical field are critical to the enhancement to be observed.

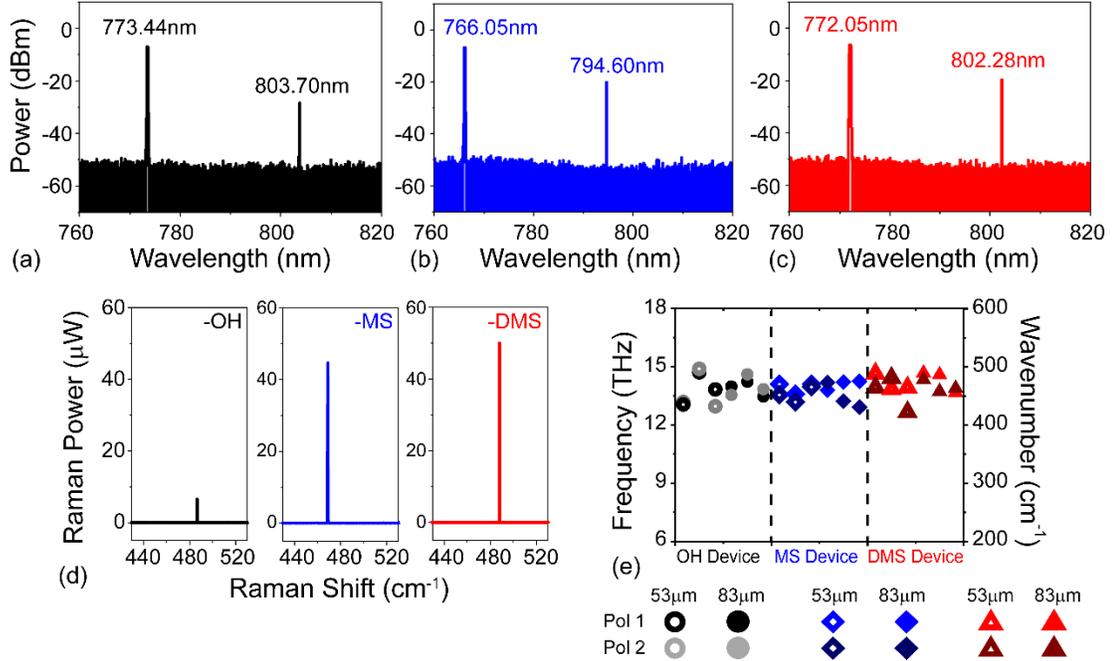

Figure 4. Representative emission spectra of (a) OH, (b) MS, and (c) DMS surface functionalized devices with ~53 μm diameters at one polarization state. All devices have similar Q factors and are pumped using ~350 μW coupled power. The pump and emissions wavelengths are indicated. (d) A comparison of the Raman peaks of the -OH, -MS and -DMS functionalized devices shown in (a), (b), and (c), respectively. Both MS and DMS functionalized devices show strong Raman lasing peaks while the OH functionalized device shows weaker emission. The Raman shifts for all devices fall within the expected range for the Si-O-Si bond (420 cm$^{-1}$ to 500 cm$^{-1}$). (e) All Raman shifts measured for all devices tested with two different polarization states. Two different device sizes (~53 μm (hollow symbol) and ~83 μm (solid symbol)) are measured. All values fall within the Si-O-Si vibrational band.

To more quantitatively analyse the enhancement in Raman lasing due to the SSRS, it is necessary to measure the lasing efficiency and lasing threshold. Figure 5 presents Raman lasing threshold curves for all three types of devices excited using two different optical mode polarizations. Figure 5 (a) shows the Raman lasing behaviours from two different polarization states in the OH functionalized silica devices. The similarity in slope and threshold values between the two data sets demonstrates that the polarization state of the incident field has little effect on the performance when a OH surface terminated device is used. This result is not surprising given the amorphous structure of OH device which consists of randomly distributed Si-O-Si vibrational modes resulting in polarization independent behaviour. This behaviour has also been observed in previous work with amorphous or randomly ordered materials[5,29].

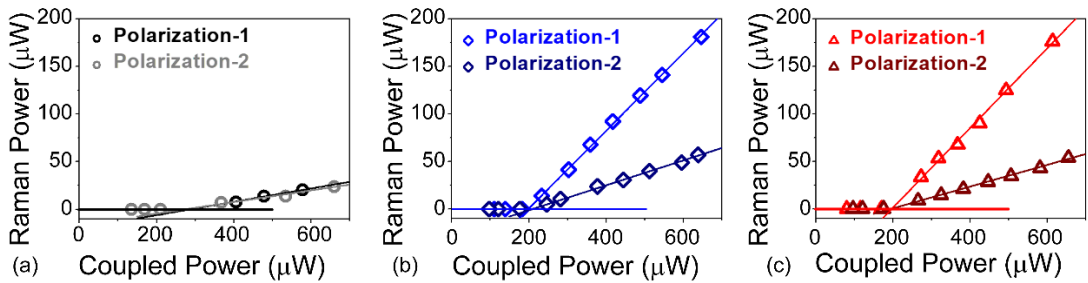

Figure 5. Polarization dependent surface Raman lasing behaviours from one (a) OH (b) MS, and (c) DMS functionalized device with 53μm diameter excited at two different polarizations. There was no difference observed for the amorphous OH device, but a strong dependence on lasing efficiency was observed for the MS and DMS functionalized devices. All data is in Table S2 and S3.

In contrast, -MS and -DMS devices exhibit a significant performance improvement as compared to the -OH devices, and they show a strong dependence on the polarization of the input optical light (Figure 5 (b) and (c)). In both -MS and -DMS devices, the Raman lasing efficiency of one polarization state is substantially higher than that of the other polarization state (~3x), and both values are significantly increased compared with uncoated silica devices (maximum of ~ 6x). As discussed, the improvement in polarization dependent Raman lasing performance is attributed to the highly oriented, tilted surface Si-O-Si Raman mode. The efficiency results for all devices are shown in Figure 6(a), and this enhancement is consistently observed.

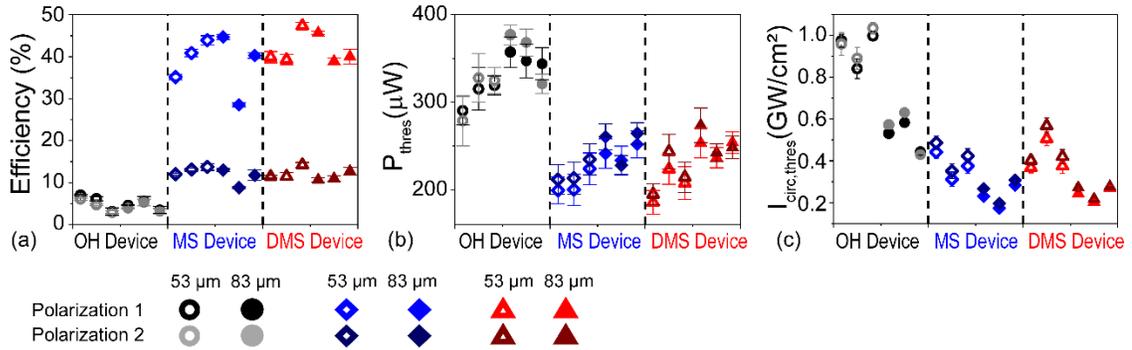

Figure 6. Comparison of Raman lasing performance from OH, MS, and DMS devices with two different diameters excited at two different polarizations: ~ 53 μm (hollow symbols) and ~ 83 μm (solid symbols). (a) The Raman lasing efficiency for the -OH functionalized device was independent of all parameters, but the MS and DMS devices exhibited a strong dependence on polarization. Additionally, the expected 3x difference between polarization states for a given surface chemistry is observed. (b) While the Raman lasing threshold power for all devices exhibited a slight dependence on diameter, there was a pronounced dependence on surface chemistry (~1.5x). (c) When threshold power is converted to circulating intensity, the expected 2x geometric dependence within a given device type becomes clearly evident. Additionally, by removing the mode area and Q variation, the expected ~2x decrease in Raman lasing threshold due to the MS and DMS layers is revealed.

Using the results presented in Figure 6(a), the relative enhancement for each device diameter and each surface functionalization can be calculated and compared to the expected value of $\cos^2(30°)/\cos^2(60°) = 3$. For the MS devices, the ratios are $3.17 \pm 0.26$ and $3.53 \pm 0.82$ for the 53 μm and 83 μm devices, respectively. For the DMS devices, the ratios are $3.50 \pm 0.44$ and 3.73

± 0.35 for the 53 µm and 83 µm devices, respectively. Given the error in the measurement, these findings are in very good agreement with the predicted value of 3. Additionally, as further confirmation, the efficiency is independent of device geometry.

Figure 6 (b) summarizes the threshold power ($P_{thres}$) required to achieve stimulated Raman scattering in all of the devices. Overall, the MS and DMS devices have lower threshold powers, by about 1.5x, than that of the OH devices. In addition, for MS and DMS devices, the two polarizations have different thresholds, with higher efficiency devices achieving lower lasing thresholds in general. However, unlike the lasing efficiency, the lasing threshold exhibits a strong dependence on device geometry and Q. Therefore, when multiple devices are being compared, a more appropriate metric to consider is the optical intensity at threshold ($I_{circ,thres}$). This metric is calculated by converting $P_{thres}$ to circulating intensity using the following expression: $I_{circ,thres}=P_{circ}/A_m$ where $P_{circ}$ is the circulating power at threshold and $A_m$ is the optical mode area. Both mode area and circulating power are dependent on the major and minor radii of the device, and circulating power is dependent on quality factor. Therefore, this calculation removes the effect of optical mode variations and device to device variations. For example, using the geometries from the present experiments and assuming an input power of 1mW, the circulating intensity in the smaller devices (~53 µm) is ~ 3.96 GW/cm$^2$ whereas the circulating intensity in the larger devices (~83 µm) is ~ 1.76 GW/cm$^2$. Therefore, it would be expected that the threshold in the smaller devices would be approximately twice the threshold in the larger devices, when presented in terms of the circulating intensity in the device. This trend is clearly observed across all data for the same device types in Figure 6 (c). Moreover, further improvements in device performance are observed when the devices are functionalized with the DMS and MS layers. Notably, the thresholds decrease by an additional ~2x.

Conclusion

In conclusion, surface stimulated Raman lasing is demonstrated by grafting a highly oriented organic monolayer on an integrated ultra-high-Q optical microcavity. The ordered asymmetric monolayers of MS and DMS are uniform on the surface of the toroidal microresonators, allowing $Q > 10^7$ to be maintained. Using a 765nm excitation source, both the MS and DMS devices exhibit surface Raman lasing efficiencies above 40% or 10x improvement over non-functionalized devices. Moreover, the Raman signal exhibits polarization dependent behaviour, and the efficiency improves 3x between the two polarization states for a single surface functionalization, further confirming that the Raman emissions originate from the monolayer. Lastly, the lasing threshold is also reduced by 2x due to the monolayer. This work paves the way for probing nanoscale properties of optical molecules with on-chip devices[46–48] and for investigating quantum light-matter interactions[49–50].

Acknowledgements

The authors thank Yunfeng Xiao and Jinhui Chen for helpful discussions and Mark Veksler for scientific visualization. The authors would like to acknowledge IARPA [2016-16070100002] and the Office of Naval Research [N00014-17-2270].

Author contributions

X. S. and H. C contributed equally to this work. X. S. and H.C. conceived the project. X. S., H. C., and A. M. A. designed the experiments. H. C. fabricated the devices. X. S. functionalized the devices. X. S and H. C. conducted testing and data analysis. W.Z. conducted DFT simulation of the model molecules. D. C. performed the FEM simulation. X. S., H. C., and A. M. A. wrote the manuscript. All authors revised and commented on the manuscript. All authors have given approval to the final version of the manuscript and supplementary information.

The authors declare no competing or conflicts of interest. Correspondence and requests for materials should be addressed to Andrea Armani (armani@usc.edu) or Xiaoqin Shen (shenxq@shanghaitech.edu.cn).

Data Availability Statement

The data that support the plots with this paper and other findings of this study are available from the corresponding author upon reasonable request.

*Supplementary Information for:*

# Raman laser from an optical resonator with a grafted single molecule monolayer

*Xiaoqin Shen[1, *, ‡], Hyungwoo Choi[1, *], Dongyu Chen[2], Wei Zhao[3], and Andrea M. Armani[1,2]*

**Contents:**

1. Device Fabrication and Surface Functionalization
2. Surface Chemistry Characterization
3. Computational Modeling
4. Optical Device Characterization
5. Comparison of Raman spectra between the aligned Si-O-Si mode and disordered C-H mode in MS devices
6. Calculation of circulating power and intensity
7. References

# 1 Device Fabrication and Surface Functionalization

## 1.1 Fabrication of bare silica toroid

Silica toroidal microresonators are fabricated using three main steps: photolithography to define silica circles on silicon wafers, $XeF_2$ etching to remove silicon isotropically to obtain a suspended silica disks on silicon support pillars, and $CO_2$ laser reflow to produce silica toroids[1]. Each step will be discussed in more detail.

Intrinsic silicon wafers with a 2 µm layer of thermally grown $SiO_2$ (WRS Materials) are cleaned using acetone, methanol, and isopropanol, then dried with nitrogen gas air gun and placed on a hot plate at 120 °C for 2 minutes to remove any possible residual solvent. Hexamethyldisilazane (HMDS) is applied to the wafers to improve adhesion between the photoresist and the silica surface. The S1813 photoresist is spin-coated onto the wafer at 500 rpm for 5 seconds and 3000 rpm for 45 seconds, followed by soft bake at 95 °C for 2 minutes in order to harden the photoresist. Circles are patterned in the photoresist using UV intensity of 80 mJ/cm$^2$. The resist is developed using MF-321 developer for 1 minute, followed by a hard bake at 120 °C for 2 minutes. The pattern is transferred into the silica using buffered oxide etchant (BOE). BOE etching takes approximately 20 minutes to etch the 2 µm thickness of the thermally grown silica. The photolithography process is completed by removing the remaining photoresist completely using another acetone, methanol, and isopropanol rinsing cycle followed by drying with a nitrogen gas air gun.

Once the photolithography steps are finished, the silicon underneath the silica circle pattern is etched isotropically using a $XeF_2$ pulsed gas etcher, producing silica micro-disks which are elevated above the silicon substrate on pillars. The toroid fabrication process is finished by reflowing the as-prepared silica microdisks using a $CO_2$ laser operating at 10.6 µm. Silica has high absorption at the wavelength of $CO_2$ laser, causing the silica disk to be reflowed into a silica toroid.

**1.2  Surface Chemistry**

The prepared silica toroidal microcavities are treated by $O_2$ plasma using an SCE 104 plasma system (Anatech USA) to generate hydroxyl groups on the surface. Organic chlorosilane agents (methyl trichlorosilane or dimethyl dichlorosilane) are deposited on the surface of the silica resonators using chemical vapor deposition at room temperature for about 7 min. An overview of this process and the chemical layers formed is shown in Figure S1.

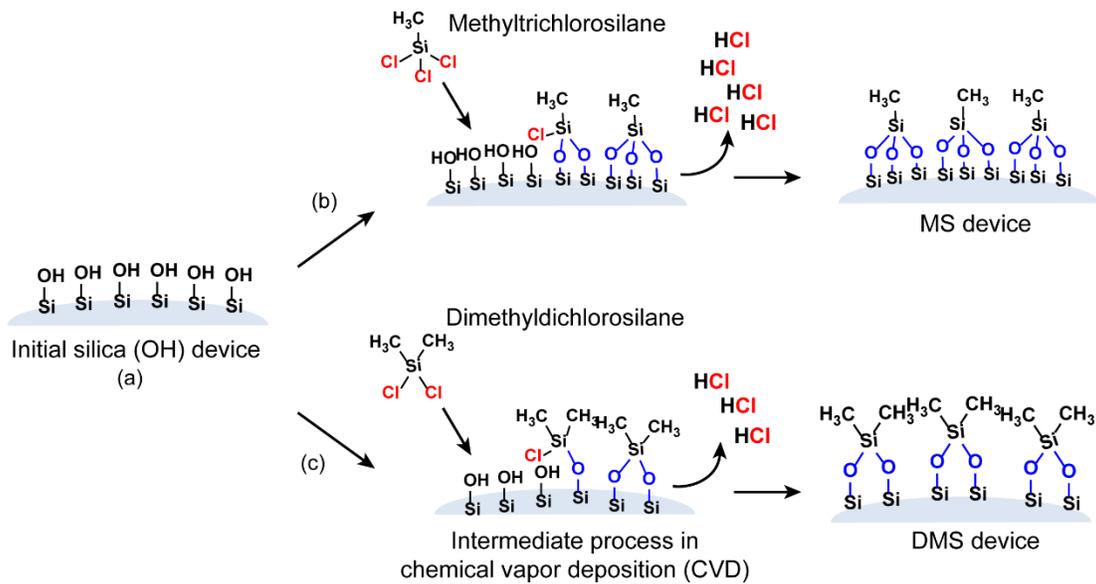

Figure S1. Schematic of surface chemistry procedure used to create the oriented monolayers of MS and DMS on the device surfaces. (a) The initial device intrinsically has a layer of hydroxyl groups. (b)/(c) Using vapor deposition, either methyl trichlorosilane (b) or dimethyl dichlorosilane (c) are used to silanate the device surfaces, forming the molecular monolayers of MS and DMS, respectively.

During the deposition process, the free chlorosilane molecules in the chemical vapor inside a closed chamber diffuse and contact the surface of the silica resonators. These highly reactive chlorosilane molecules react with the hydroxyl groups on the device surface, yielding new Si-O-Si bonds and releasing hydrochloride molecules[2,3]. Because the reactive Si-Cl sites in the asymmetric chlorosilane molecules can only react with the hydroxyl groups on the surface, the reaction spontaneously forms a single, oriented surface molecular layer. Note, this reaction is intrinsically self-limiting, allowing only a single monolayer to form. This process yields uniform grafted MS or DMS oriented mono-molecular layers on the surface of the resonant cavity devices.

Assuming the initial Si-O bonds are vertical to the surface, and because the theoretical O-Si-O bond angle is ~110° (taken from reference[4,5]), the Si-O-Si bond angle is estimated to be from ~120° to ~125°. The new generated Si-O-Si bonds are constrained and fixed on the surface, and the estimated Si-O-Si bond angle is consistent with the result in previous report[4].

## 2  Surface Chemistry Characterization

### 2.1  Raman spectroscopy

The spontaneous Raman spectra measurements are performed using a Reinsaw InVia micro-Raman spectrometer with a 100x objective lens. The Raman spectroscopy is performed on three samples to mimic the chemical structures. The first sample is silica (a glass slide). The second and third samples are MS and DMS thin films, prepared by drop-casting methyltrichlorosilane and dimethyldichlorosilane on silicon wafers followed by exposure to water vapor for 30 min and then vacuum drying.

This sample preparation approach results in a disordered thin film which is expected to have reduced gain as compared to an oriented layer. However, the comparison between sample types should prove informative. The results are presented in Figure S2. For both MS and DMS thin films, the C-H vibration of the methyl groups shows intensive characteristic Raman peaks at ~ 2980 cm$^{-1}$, and the Si-O-Si vibration shows Raman peaks at about 460 cm$^{-1}$. Because the Raman peaks of interest in this study are from Si-O-Si vibration, the Raman spectra in the range of ~350 cm-1 to 550 cm$^{-1}$ from the MS, DMS and silica thin films are directly compared. Comparative Raman gain coefficient at about 460 cm$^{-1}$ of MS and DMS thin film are calculated using the silica sample as a reference.

The Raman gain coefficient, $g_R$, is related to the spontaneous Raman scattering cross section as described in literature[6].

$$g_R = \frac{\sigma \lambda_S^3}{c^2 h \varepsilon (n+1)}$$

, where σ is the spontaneous Raman cross section, $\lambda_S$ is the stokes wavelength, $h$ is Planck's constant, $\varepsilon$ is the dielectric constant, and n is the Bose-Einstein population factor. The Raman gain coefficient of silica ($g_R$= 0.53 × 10$^{-13}$ m/W) was taken from literature[7].

Based on the data, the gain coefficient values for the MS and DMS layers are estimated to be ~13.5 × 10$^{-13}$ m/W and ~12.9 × 10$^{-13}$ m/W. The gain coefficient values for the C-H Raman modes are also calculated to be ~ 1.62 × 10$^{-11}$ m/W and 2.97 × 10$^{-11}$ m/W for MS and DMS layers, respectively. These values are ~12 times and ~23 times higher than the values for the Si-O Raman modes of the MS and DMS layers, respectively. However, once ordered monolayers are formed using the previously described surface chemistry, the Si-O-Si Raman modes will be anchored to the surface and aligned parallel to the direction of the electrical field while the C-H Raman modes will continue to be randomly distributed and disordered. The alignment should greatly increase the signal generated.

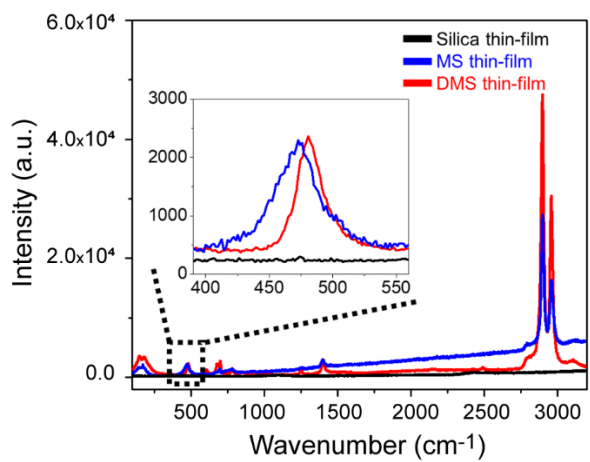

Figure S2. Spontaneous Raman spectra of silica, MS and DMS thin films taken using a reflective micro-Raman spectrometer. Inset: The area of interest corresponding to the Si-O-Si vibration in the range from 390 cm$^{-1}$ to 560 cm$^{-1}$.

## 2.2   X-ray Photoelectron spectroscopy (XPS)

The formation of the mono-molecular single layers is characterized by both indirect and direct methods. The indirect method leverages a mimic molecule with a Cl-substitution to enable XPS (X-ray Photoelectron spectroscopy) analysis, and the direct method uses Raman spectroscopy.

The MS and DMS molecular layers consist only of carbon (C) and hydrogen (H) atoms, and the devices contain silicon (Si) and oxygen (O). However, the sample chamber has an intrinsic high carbon background. Therefore, XPS cannot effectively determine the success of the surface chemistry due to the high background carbon signal. To address this limitation, the methyl silane is replaced by a chloro-substituted methyl silane. This substitution allows the same chemical vapor deposition method to be used to form a chloro-substituted methyl molecular layer on the surface. The Cl acts as a label for the methyl group, enabling the monomolecular layers to be easily distinguished from the initial bare silica sample by simply tracking the peak of binding energy of chloride in the XPS spectra. As shown in Figure S3 (a), the XPS spectrum of the chloro-substituted methyl silane molecular layers on silica shows a characteristic peak of binding energy of Cl(2p) at about 200 eV, suggesting the formation of a grafted molecular layer on the surface of silica.

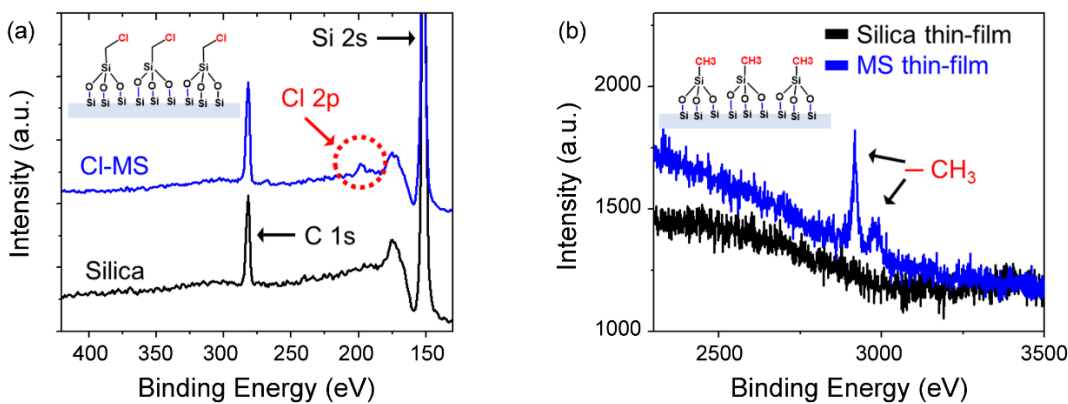

Figure S3. Characterization of the molecular monolayers. (a) Comparison of XPS spectra of a grafted chloro-substituted MS monolayer on silica and an initial bare silica. The peak binding energy at about 200 eV is the primary peak of Cl(2p) from the chloro-substituted MS molecules grafted on the surface of silica. (b) Comparison of micro-Raman spectra of a grafted MS monolayer on silica and the initial bare silica. The Raman peaks at about 2900 cm$^{-1}$ originate from the methyl group of the MS molecules grafted on the silica surface. These data confirm the formation of the grafted molecular layers on the surface of silica resonators using the same chemical vapor deposition method.

To directly characterize the materials, micro-Raman spectroscopy is used. Compared to the initial bare silica surface, the spectra of the MS or DMS coated surfaces clearly show Raman peaks at about 2900 cm$^{-1}$ (Figure S3 (b)), even though the intensity from the molecular monolayer is typical low. These peaks are the characteristic Raman peaks of C-H in methyl groups, directly confirming the formation of MS and DMS molecular layers on the surface silica.

## 2.3 Refractive index of organic molecules

Given the less than 1 nm thickness of the organic molecules layer, it is extremely challenging to accurately characterize the refractive index of the film with an ellipsometer. An alternative approach to directly measuring the film index is to measure the change in free spectral range (FSR) upon the addition of the monolayer. Because the FSR is directly related to the effective refractive index of the optical mode, this method is a more accurate approach for analysing the impact of the film on the optical mode.

In a non-functionalized silica device, the effective refractive index is 1.4448. In MS and DMS functionalized devices, the effective indices are 1.4462 and 1.4463, respectively. Thus, the surface coatings have minimal impact on the refractive index, which is expected given their material composition and thickness.

## 3 Computational Modelling

## 3.1 Density Functional Theory (DFT) of molecular hyperpolarizabilities

The theoretical vibrational second hyperpolarizabilities of three model molecules $Si(OSiH_3)_2(OH)_2$, $Si(OSiH_3)_2(OH)(CH_3)$, and $Si(OSiH_3)_2(CH_3)_2$ were computed using Gaussian 2003 software package with RHF and DFT calculations as described in previous literature[8,9]. The levels and basis sets used here were RHF/6-311+G(3df,2p). The calculations focused on the breathing mode of the compounds around 500 cm$^{-1}$. The ring breathing mode of benzene at 992 cm$^{-1}$ ($v_2$ in Herzberg notation) served as an internal standard to compare across molecules. Here, we used its Raman $\gamma^{ref}$ of 1.5 ×10$^{-35}$ esu, reported by Levenson and Bloembergen[10], for the second hyperpolarizability calculations. The simulated spectra were plotted with a Lorentzian broadening using 8 cm$^{-1}$ for the full width at half magnitude (FWHM). Data are shown in Figure S4 and Table S1.

It is important to note that the simulated spectra are for the three simplified model molecules that are not anchored to a substrate. Therefore, the results are expected to be slightly different from the experimental data of the three spun-coat deposited films or the oriented films on the device surfaces. However, the simulation results should be informative as to the relative impact of the asymmetric structure on the microscopic molecular nonlinear properties.

Frequencies were calibrated by the experimental peak position of the C-H vibration so that the peak studied for each sample was close to the experimental value of the breathing mode at ~450 cm$^{-1}$. As can be seen in Table S1, the $\gamma^{vib}$ corresponding to the Si-O-Si mode are calculated to be 1.2 and 1.7 × 10$^{-36}$ cm$^6$/erg for the MS and DMS model molecules, respectively. These values are approximately two times larger than the silicon dioxide model molecules (7.8 × 10$^{-37}$ cm$^6$/erg). This increase is attributed to the asymmetry that is introduced by the organic methyl group in the siloxane structure.

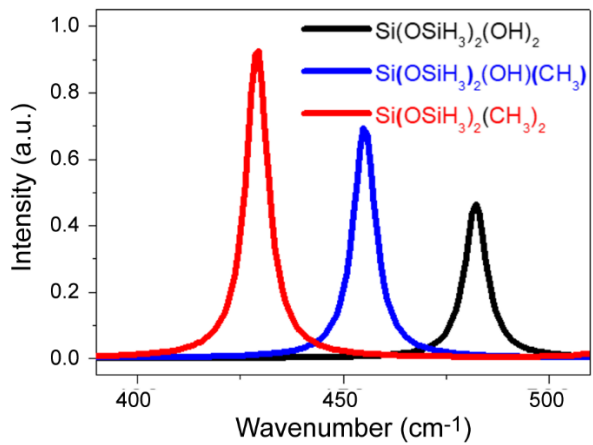

Figure S4. RHF-based, simulated Raman spectra of three model compounds, $Si(OSiH_3)_2(OH)_2$, $Si(OSiH_3)_2(OH)(CH_3)$, and $Si(OSiH_3)_2(CH_3)_2$, plotted with a Lorentzian broadening (FWHM = 8 cm$^{-1}$). Frequencies were calibrated by the experimental peak position of the C-H vibration so that the peak studied for each sample was close to the experimental value of the breathing mode at ~450 cm$^{-1}$.

**Table S1.** Raman intensity ratios $(I^S/I^{ref})_C$ and Raman γ values determined from computed Raman activities using RHF/6-311+G(3df,2p) method.

| Molecular Models | Exp. frequency (cm$^{-1}$) | $(I^S/I^{ref})_C$ | γ (cm$^6$/erg) ×10$^{-37}$ |
|---|---|---|---|
| $C_6H_6$ | 992 | 1.00 | 150 |
| $Si(OSiH_3)_2(OH)_2$ | 462 | 0.0595 | 7.8 |
| $Si(OSiH_3)_2(OH)(CH_3)$ | 468 | 0.0871 | 12 |
| $Si(OSiH_3)_2(CH_3)_2$ | 429 | 0.131 | 17 |

## 3.2 Finite element method (FEM) modelling of optical field

We use COMSOL Multiphysics finite element method (FEM) to model the optical modes of the cavity. Briefly, we draw a cross section of the optical resonator and solve Maxwell's equations by assuming an axially symmetric mode. The device geometries (major and minor radii) and material properties are defined by the devices used in the experiments. We chose the mesh size to be $\lambda/8$ to achieve an acceptable accuracy. Once the software solves for the optical modes, we draw lines on the equatorial plane of the resonator and calculate the electric field magnitude for all three components of the field (axial, azimuthal, and radial).

The cross section of the fundamental mode along with the field amplitude profile on the equatorial plane is shown in Figure 1 (b) in the main text. In the modelling, it is assumed that the effective refractive index of the microcavities is not changed by the presence of the ultra-thin (~1 nm) monolayer. (This assumption is verified, as described in a subsequent section.) Therefore, the presence of the monolayer does not change or distort the optical mode profile. Thus, it is expected that the optical cavity can directly and efficiently interact with the monolayer.

## 4 Optical Device Characterization

### 4.1 Measurement of the intrinsic quality (Q) factor

The quality (Q) factors of the devices are characterized with a tunable narrow linewidth laser operated at 765 nm (Velocity series, Newport) by coupling light into the microcavity using a tapered optical fibre waveguide. The tapered optical fibres are fabricated by slowly pulling a single mode optical fibre (F-SC, Newport) while it is heated with a hydrogen torch. The single mode of the tapered optical fibre is confirmed by an in-situ transmission measurement with an oscilloscope during fabrication. The optical microcavity is aligned with the tapered optical fibre using a 3-axis nano-positioning stage. The output from the optical fibre is sent to a 50:50 splitter, which is connected to the photodetector and the optical spectrum analyser (OSA, YOKOGAWA AQ6370C). A schematic of the set-up is shown in the main text.

The resonant wavelength is determined by scanning across a series of wavelengths. The resonant spectrum is fit to a Lorentzian, and the loaded Q factor of the device is determined with the equation: $Q = \lambda/\Delta\lambda$, where $\lambda$ is the resonant wavelength of the device and $\Delta\lambda$ is the full-width-half-maximum of the peak, as shown at Figure S5 (a). All data and images are recorded with a computer integrated with PCI GPIB, function generator, and oscilloscope. A general laser communication port (PCI GPIB) and a function generator are connected to a tunable laser, and they are used to finely tune the laser wavelength and locate the resonant wavelength of the device. During these measurements, the scan rate and range of the laser is optimized to ensure the linewidth is not distorted due to thermal or other effects.

Figure S5 (a) is one of the representative normalized transmission spectra, which is recorded from an oscilloscope connected to the photodetector. Due to the ultra-high Q of the device, the mode-splitting shown in the transmission spectra occurs in most measurements. This behaviour commonly occurs in ultra-high Q devices because the light is coupled into two different directions of optical modes (clockwise and counter-clockwise)[1].

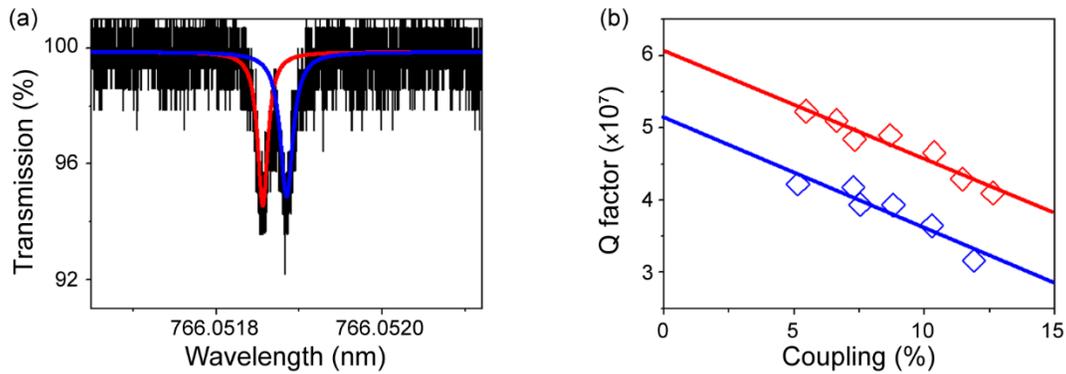

Figure S5. (a) One of the representative transmission spectra used to obtain the loaded Q factors from the MS device. The resonant wavelengths are 766.05186 and 766.05188 nm with the loaded Q factors of 5.23 and 4.22 x $10^7$, and coupling percentages during this measurement are 5.44 and 5.13 %, respectively. (b) The loaded Q factor data with respect to the coupling percentage. The y-intercept of each linearly fitted data corresponds to the intrinsic Q factor of either clockwise and counter-clockwise, which are 6.08 and 5.16 x $10^7$.

Figure S5 (b) contains the loaded Q factors as a function of the coupling percentage to calculate the intrinsic Q factor of a device. The intrinsic Q factor is determined by measuring the loaded Q from the transmission spectra over a range of coupling percentage and by removing the extrinsic losses, or in this case the coupling losses, using a coupled-cavity model.

## 4.2  Raman lasing threshold and efficiency

The OSA is used to detect the generated Raman emission that are coupled back into the tapered optical fibre. A series of OSA spectra are obtained by finely tuning the power propagating through the tapered optical fibre. At each power, the OSA spectrum, the power out of the optical fibre, and the percentage of power coupled into the optical resonator are recorded. From the OSA spectrum, the generated Raman lasing power is recoded in the units of dBm. This value is subsequently converted to µW. The power coupled into the device is calculated by multiplying the power in the tapered optical fibre with the coupling percentage into the device. Then, the Raman lasing power is plotted as a function of the coupled power to the resonators. The data is linearly fitted as the output Raman power is directly proportional to the coupled power. In the graph, slope of the line indicates the Raman lasing efficiency while the x-intercept of the line corresponds to the Raman lasing threshold.

## 5  Comparison of Raman spectra between the aligned Si-O-Si mode and disordered C-H mode in MS devices.

It is worth noting that emission peaks around 980 nm (Raman shift in the range of ~2850 $cm^{-1}$ to ~2980 $cm^{-1}$ with 765 nm pump) from the C-H stretching modes of the methyl groups on the functionalized devices are not observed, even under high input power, as shown at Figure S6. In contrast to the oriented Si-O-Si vibrational mode that is anchored directly on the surface and highly aligned with direction of the electric field, the orientation of the C-H stretching modes of the methyl groups in the surface molecules are randomly distributed and are not parallel to the polarization direction of the electrical field. The absence of emission peaks at ~980 nm provides agrees with that the orientation of the sub-molecular vibrational mode is critical to the enhancement observed.

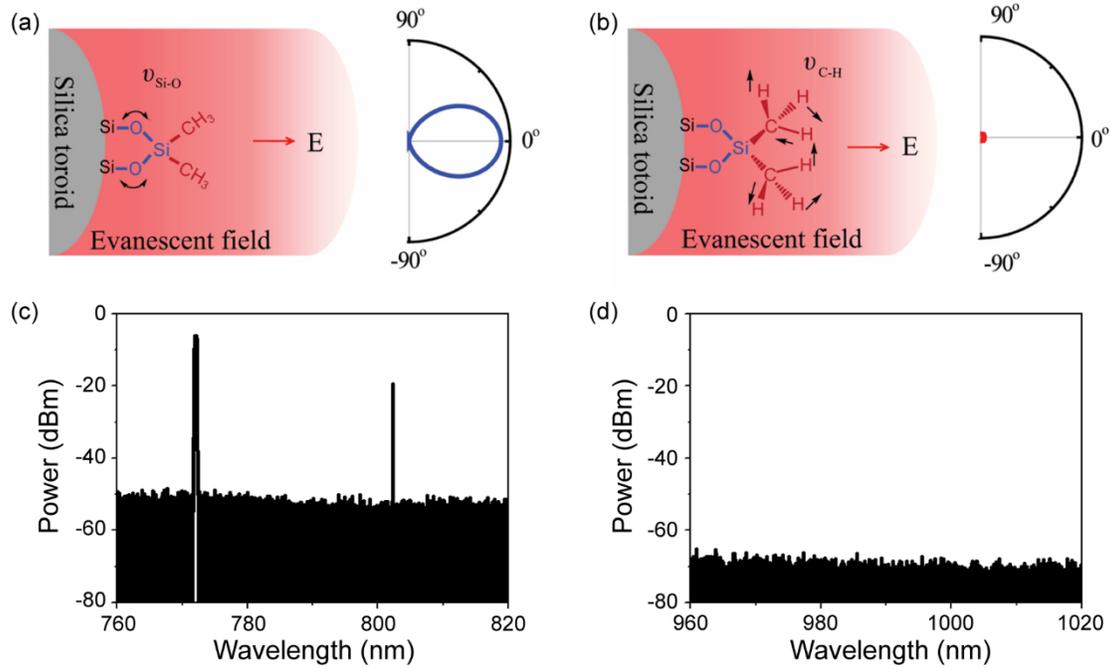

Figure S6. (a) Schematic of the aligned Si-O mode of the surface MS layer (Left). The mode direction is parallel to the direction electrical field, resulting in maximum Raman efficiency (Right). (b) Schematic of the disordered C-H mode of the surface MS layer (Left). The mode directions, both in-plane and out-plane (not shown in the schematic), are randomly distributed and are not parallel to the direction transverse electric field. The Raman efficiency is not amplified and is not dependent on the propagation direction of the TE mode. (c) Output spectrum (760 nm to 820 nm) of a DMS device (50 μm; Q = 5.4 × $10^7$) pumped at ~765 nm with coupled power of about 400 μW. The Raman lasing peak at about 793 nm corresponding to the Si-O-Si mode. (d) Output spectrum (960 nm to 1020 nm) of the same MS device as in part (c) also pumped at ~765 nm. However, the coupled power is increased to about 1mW. Even at the higher power, no Raman peak is observed at about 980 nm corresponding to the C-H mode (~ 2900 cm$^{-1}$).

## 6 Calculation of circulating power and intensity

To compare to performance of different devices, the threshold power needed for Raman lasing is calculated. This calculation begins by determining the circulating power inside the device ($P_{circ}$) using:

$$P_{circ} = P_{in} Q_0 \frac{\lambda}{\pi^2 nR} \frac{K}{(1+K)^2} \qquad \text{Equation S1.}$$

where $\lambda$ is the resonance wavelength, R is the device principle radius, $P_{in}$ in the input power, $Q_0$ is the intrinsic quality factor of the device, $n$ is the effective refractive index, and $K$ is the coupling coefficient, which is directly related with the transmission of the input light at the resonance wavelength as follows:

$$T = \frac{(1-K)^2}{(1+K)^2} \qquad \text{Equation S2.}$$

The peak intensity circulating into the cavity is presented as $P_{circ}/A_m$, where $A_m$ is the optical mode area. The value of $P_{circ}/A_m$ has taken the variations in Q in different devices into account; thus, it is feasible to direct compare the Raman thresholds in different devices. The mode area is defined the below equation, and the value is calculated using the previously described FEM simulations:

$$A_m = \frac{\int \varepsilon(r)|E|^2 dA}{\max(\varepsilon(r)|E|^2)} \qquad \text{Equation S3.}$$

Table S2. Summary of 53 μm diameter devices.

| | OH | | OH | | OH | | MS | | MS | | MS | | DMS | | DMS | | DMS | |
|---|---|---|---|---|---|---|---|---|---|---|---|---|---|---|---|---|---|---|
| | Pol.1 | Pol.2 | Pol.1 | Pol.2 | Pol.1 | Pol.2 | Pol.1 | Pol.2 | Pol.1 | Pol.2 | Pol.1 | Pol.2 | Pol.1 | Pol.2 | Pol.1 | Pol.2 | Pol.1 | Pol.2 |
| Q (x $10^7$) | 8.80 ± 0.17 | 9.02 ± 0.24 | 6.99 ± 0.50 | 7.12 ± 0.25 | 8.19 ± 0.90 | 8.35 ± 0.34 | 5.81 ± 0.66 | 6.02 ± 0.26 | 4.09 ± 0.90 | 4.31 ± 0.37 | 4.39 ± 1.07 | 4.73 ± 0.81 | 5.20 ± 1.33 | 5.41 ± 0.29 | 5.93 ± 1.27 | 6.11 ± 0.98 | 4.74 ± 0.82 | 5.14 ± 0.59 |
| Shift ($cm^{-1}$) | 435.41 | 439.63 | 490.78 | 496.67 | 461.40 | 432.29 | 470.21 | 451.97 | 453.34 | 439.63 | 470.55 | 465.61 | 488.74 | 464.98 | 462.24 | 481.01 | 464.62 | 422.29 |
| Efficiency (%) | 7.02 ± 0.17 | 6.1 ± 0.24 | 6.18 ± 0.50 | 4.76 ± 0.25 | 3.10 ± 0.90 | 2.95 ± 0.34 | 35.16 ± 0.66 | 12.02 ± 0.26 | 40.90 ± 0.90 | 13.11 ± 0.37 | 43.91 ± 1.07 | 13.73 ± 0.81 | 39.89 ± 1.33 | 11.43 ± 0.29 | 39.28 ± 1.27 | 11.47 ± 0.98 | 47.36 ± 0.82 | 14.21 ± 0.59 |
| Threshold (μW) | 290.35 ± 16.30 | 278.69 ± 28.73 | 315.23 ± 24.35 | 327.69 ± 27.81 | 319.27 ± 10.92 | 324.81 ± 14.71 | 199.15 ± 15.16 | 211.65 ± 16.79 | 199.90 ± 17.91 | 213.11 ± 18.37 | 223.91 ± 18.07 | 234.71 ± 17.81 | 185.36 ± 13.63 | 194.79 ± 11.80 | 223.28 ± 17.18 | 243.54 ± 19.84 | 207.36 ± 18.82 | 214.21 ± 17.59 |
| Threshold ($GW/cm^2$) | 0.975 ± 0.031 | 0.959 ± 0.055 | 0.841 ± 0.047 | 0.891 ± 0.053 | 0.998 ± 0.021 | 1.035 ± 0.028 | 0.442 ± 0.029 | 0.486 ± 0.032 | 0.312 ± 0.034 | 0.351 ± 0.035 | 0.375 ± 0.034 | 0.424 ± 0.034 | 0.368 ± 0.026 | 0.402 ± 0.023 | 0.505 ± 0.033 | 0.568 ± 0.038 | 0.375 ± 0.036 | 0.420 ± 0.034 |

Table S3. Summary of 83 μm diameter devices.

| | OH | | OH | | OH | | MS | | MS | | MS | | DMS | | DMS | | DMS | |
|---|---|---|---|---|---|---|---|---|---|---|---|---|---|---|---|---|---|---|
| | Pol.1 | Pol.2 | Pol.1 | Pol.2 | Pol.1 | Pol.2 | Pol.1 | Pol.2 | Pol.1 | Pol.2 | Pol.1 | Pol.2 | Pol.1 | Pol.2 | Pol.1 | Pol.2 | Pol.1 | Pol.2 |
| Q (x $10^7$) | 8.47 ± 0.14 | 8.64 ± 0.09 | 9.56 ± 1.10 | 9.75 ± 0.76 | 7.35 ± 0.84 | 7.62 ± 0.67 | 5.44 ± 0.60 | 5.82 ± 0.38 | 4.30 ± 0.49 | 4.88 ± 0.20 | 6.40 ± 0.66 | 6.62 ± 1.23 | 5.46 ± 0.41 | 5.66 ± 0.49 | 4.89 ± 0.84 | 5.11 ± 0.67 | 6.21 ± 1.74 | 6.35 ± 0.92 |
| Shift ($cm^{-1}$) | 466.35 | 451.84 | 474.55 | 487.23 | 449.23 | 461.82 | 459.97 | 473.19 | 474.00 | 441.26 | 475.31 | 430.84 | 489.46 | 478.15 | 486.51 | 457.91 | 455.83 | 463.17 |
| Efficiency (%) | 4.65 ± 0.14 | 3.92 ± 0.09 | 5.60 ± 1.10 | 5.31 ± 0.76 | 3.51 ± 0.84 | 3.17 ± 0.67 | 44.72 ± 0.60 | 12.99 ± 0.38 | 28.53 ± 0.49 | 8.80 ± 0.20 | 40.30 ± 0.66 | 11.75 ± 1.23 | 45.66 ± 0.41 | 10.66 ± 0.49 | 38.90 ± 0.84 | 10.97 ± 0.67 | 39.96 ± 1.74 | 12.70 ± 0.92 |
| Threshold (W) | 357.29 ± 17.48 | 377.08 ± 11.34 | 347.23 ± 19.43 | 368.36 ± 15.42 | 343.88 ± 18.33 | 320.92 ± 11.19 | 241.11 ± 16.88 | 260.53 ± 14.86 | 233.50 ± 16.30 | 227.8 ± 11.12 | 251.75 ± 15.66 | 264.30 ± 12.23 | 253.28 ± 17.18 | 273.53 ± 19.84 | 236.44 ± 11.54 | 242.34 ± 10.28 | 254.23 ± 12.33 | 248.42 ± 15.84 |
| Threshold ($GW/cm^2$) | 0.532 ± 0.015 | 0.572 ± 0.010 | 0.583 ± 0.017 | 0.631 ± 0.013 | 0.444 ± 0.016 | 0.430 ± 0.010 | 0.230 ± 0.015 | 0.266 ± 0.013 | 0.174 ± 0.014 | 0.195 ± 0.010 | 0.283 ± 0.014 | 0.307 ± 0.011 | 0.243 ± 0.015 | 0.272 ± 0.017 | 0.203 ± 0.010 | 0.218 ± 0.010 | 0.270 ± 0.011 | 0.277 ± 0.014 |